\documentclass[twocolumn]{aastex63}

\usepackage{float}              
\usepackage[utf8]{inputenc}
\usepackage{gensymb}
\usepackage[encapsulated]{CJK}
\usepackage{ucs}
\usepackage{devanagari}

\newcommand{\uplimmax}{13.5}
\newcommand{\uplimmin}{0.4}
\newcommand{\dipoleamplitude}{124}
\newcommand{\dipoleerror}{4}

\newcommand{\binwidth}{12}
\newcommand{\pvalue}{$0.54$ }

\newcommand{\cntext}[1]{\begin{CJK}{UTF8}{gbsn}#1\end{CJK}}

\def\EPSFIG[#1]#2#3#4#5{          
\begin{figure}[#5]               
\begin{center}                  %
\includegraphics[#1]{#2}        %
\end{center}                    %
\caption{#3}                    %
\label{#4}                      %
\end{figure}                    %
}

\begin{document}

\title{Two-year Cosmology Large Angular Scale Surveyor (CLASS) Observations: A Measurement of Circular Polarization at 40\,GHz}
\author[0000-0002-0024-2662]{Ivan~L. Padilla}
\affiliation{Department of Physics and Astronomy, Johns Hopkins University,
3701 San Martin Drive, Baltimore, MD 21218, USA}
\correspondingauthor{Ivan L.~Padilla}
\email{padilla@jhu.edu}

\author[0000-0001-6976-180X]{Joseph~R. Eimer}
\affiliation{Department of Physics and Astronomy, Johns Hopkins University,
3701 San Martin Drive, Baltimore, MD 21218, USA}

\author[0000-0002-4820-1122]{Yunyang Li (\cntext{李云炀}$\!\!$)}
\affiliation{Department of Physics and Astronomy, Johns Hopkins University, 
3701 San Martin Drive, Baltimore, MD 21218, USA}

\author{Graeme~E. Addison}
\affiliation{Department of Physics and Astronomy, Johns Hopkins University, 
3701 San Martin Drive, Baltimore, MD 21218, USA}

\author{Aamir Ali}
\affiliation{Department of Physics,
University Of California,
Berkeley, CA 94720, USA}
\affiliation{Department of Physics and Astronomy, Johns Hopkins University,
3701 San Martin Drive, Baltimore, MD 21218, USA}

\author[0000-0002-8412-630X]{John~W. Appel}
\affiliation{Department of Physics and Astronomy, Johns Hopkins University, 
3701 San Martin Drive, Baltimore, MD 21218, USA}

\author[0000-0001-8839-7206]{Charles~L.Bennett}
\affiliation{Department of Physics and Astronomy, Johns Hopkins University, 
3701 San Martin Drive, Baltimore, MD 21218, USA}

\author[0000-0001-8468-9391]{Ricardo Bustos}
\affiliation{Facultad de Ingenier\'ia, Universidad Cat\'olica de la Sant\'isima Concepci\'on, Alonso de Ribera
2850, Concepci\'on, Chile}

\author{Michael K. Brewer}
\affiliation{Department of Physics and Astronomy, Johns Hopkins University,
3701 San Martin Drive, Baltimore, MD 21218, USA}

\author[0000-0003-1127-0965]{Manwei Chan}
\affiliation{Department of Physics and Astronomy, Johns Hopkins University,
3701 San Martin Drive, Baltimore, MD 21218, USA}
\affiliation{Department of Aeronautics and Astronautics, Massachusetts Institute of Technology, 77 Massachusetts Ave, Cambridge, MA 02139, USA}

\author[0000-0003-0016-0533]{David~T. Chuss}
\affiliation{Department of Physics, Villanova University, 800 Lancaster Avenue, Villanova, PA 19085, USA
}
\author{Joseph Cleary}
\affiliation{Department of Physics and Astronomy, Johns Hopkins University,
3701 San Martin Drive, Baltimore, MD 21218, USA}

\author[0000-0002-0552-3754]{Jullianna Couto}
\affiliation{Department of Physics and Astronomy, Johns Hopkins University,
3701 San Martin Drive, Baltimore, MD 21218, USA}

\author[0000-0002-1708-5464]{Sumit Dahal ({\dn \7{s}Emt dAhAl}) } %
\affiliation{Department of Physics and Astronomy, Johns Hopkins University,
3701 San Martin Drive, Baltimore, MD 21218, USA}

\author{Kevin Denis}
\affiliation{Goddard Space Flight Center, 8800 Greenbelt Road, Greenbelt, MD 20771, USA}

\author{Rolando D\"unner}
\affiliation{Instituto de Astrof\'isica and Centro de Astro-Ingenier\'ia, Facultad de F\'isica, Pontificia Universidad Cat\'olica de Chile, Av. Vicu\~na Mackenna 4860, 7820436 Macul, Santiago, Chile}

\author[0000-0002-4782-3851]{Thomas~Essinger-Hileman}
\affiliation{Goddard Space Flight Center, 8800 Greenbelt Road, Greenbelt, MD 20771, USA}

\author[0000-0002-2061-0063]{ Pedro Flux\'a }
\affiliation{Instituto de Astrof\'isica and Centro de Astro-Ingenier\'ia, Facultad de F\'isica, Pontificia Universidad Cat\'olica de Chile, Av. Vicu\~na Mackenna 4860, 7820436 Macul, Santiago, Chile}

\author[0000-0001-6519-502X]{Saianeesh K. Haridas } %
\affiliation{Department of Physics and Astronomy, Johns Hopkins University,
3701 San Martin Drive, Baltimore, MD 21218, USA}

\author[0000-0003-1248-9563]{Kathleen Harrington}
\affiliation{Department of Physics and Astronomy, Johns Hopkins University,
3701 San Martin Drive, Baltimore, MD 21218, USA}
\affiliation{Department of Physics, University of Michigan, Ann Arbor, MI, 48109, USA}

\author{Jeffrey Iuliano}
\affiliation{Department of Physics and Astronomy, Johns Hopkins University,
3701 San Martin Drive, Baltimore, MD 21218, USA}

\author{John Karakla}
\affiliation{Department of Physics and Astronomy, Johns Hopkins University,
3701 San Martin Drive, Baltimore, MD 21218, USA}

\author[0000-0003-4496-6520]{Tobias~A. Marriage}
\affiliation{Department of Physics and Astronomy, Johns Hopkins University, 
3701 San Martin Drive, Baltimore, MD 21218, USA}

\author{Nathan J.~Miller}
\affiliation{Department of Physics and Astronomy, Johns Hopkins University, 
3701 San Martin Drive, Baltimore, MD 21218, USA}
\affiliation{Goddard Space Flight Center, 8800 Greenbelt Road, Greenbelt, MD 20771, USA}

\author{Carolina N\'u\~nez}
\affiliation{Department of Physics and Astronomy, Johns Hopkins University,
3701 San Martin Drive, Baltimore, MD 21218, USA}

\author{Lucas Parker}
\affiliation{Space and Remote Sensing, MS DB244, Los Alamos National Laboratory,
Los Alamos, NM 87544, USA}
\affiliation{Department of Physics and Astronomy, Johns Hopkins University, 3701 San Martin Drive, Baltimore, MD 21218, USA}

\author[0000-0002-4436-4215]{Matthew A. Petroff}
\affiliation{Department of Physics and Astronomy, Johns Hopkins University,
3701 San Martin Drive, Baltimore, MD 21218, USA}

\author{Rodrigo Reeves}
\affiliation{CePIA, Departamento de Astronom\'ia, Universidad de Concepci\'on, Concepci\'on, Chile}

\author[0000-0003-4189-0700 ]{Karwan Rostem}
\affiliation{Goddard Space Flight Center, 8800 Greenbelt Road, Greenbelt, MD 20771, USA}

\author{Robert W. Stevens}
\affiliation{Quantum Sensors Group, National Institute of Standards and Technology, 325 Broadway, Boulder, CO 80305, USA}

\author{Deniz Augusto Nunes Valle}
\affiliation{Department of Physics and Astronomy, Johns Hopkins University,
3701 San Martin Drive, Baltimore, MD 21218, USA}

\author[0000-0002-5437-6121]{Duncan~J. Watts}
\affiliation{Department of Physics and Astronomy, Johns Hopkins University, 3701 San Martin Drive, Baltimore, MD 21218, USA}

\author{Janet~L. Weiland}
\affiliation{Department of Physics and Astronomy, Johns Hopkins University, 3701 San Martin Drive, Baltimore, MD 21218, USA}

\author[0000-0002-7567-4451]{Edward~J. Wollack}
\affiliation{Goddard Space Flight Center, 8800 Greenbelt Road, Greenbelt, MD 20771, USA}

\author[0000-0001-5112-2567]{Zhilei Xu (\cntext{徐智磊}$\!\!$)}
\affiliation{Department of Physics and Astronomy, University of Pennsylvania, 
209 South 33rd Street, Philadelphia, PA 19104, USA}
\affiliation{Department of Physics and Astronomy, Johns Hopkins University, 
3701 San Martin Drive, Baltimore, MD 21218, USA}

\date{October 2019}

\begin{abstract}
We report circular polarization measurements from the first two years of observation with the 40\,GHz polarimeter of the Cosmology Large Angular Scale Surveyor (CLASS). CLASS is conducting a multi-frequency survey covering 75\% of the sky from the Atacama Desert designed to measure the cosmic microwave background (CMB) linear E and B polarization on angular scales $1^\circ \lesssim \theta \leq 90^\circ$, corresponding to a multipole range of  $2 \leq \ell \lesssim 200$. The modulation technology enabling measurements of linear polarization at the largest angular scales from the ground, the Variable-delay Polarization Modulator, is uniquely designed to provide explicit sensitivity to circular polarization (Stokes $V$). We present a first detection of circularly polarized atmospheric emission at 40\,GHz that is well described by a dipole with an amplitude of $\dipoleamplitude\pm\dipoleerror\,\mathrm{\mu K}$ when observed at an elevation of $45^\circ$, and discuss its potential impact as a foreground to CMB experiments. Filtering the atmospheric component, CLASS places a 95\% C.L. upper limit of $\uplimmin\,\mathrm{\mu K}^2$ to $\uplimmax\,\mathrm{\mu K}^2$ on $\ell(\ell+1)C_\ell^{VV}/(2\pi)$ between $1 \leq \ell \leq 120$, representing a two-orders-of-magnitude improvement over previous limits.

\end{abstract}

\keywords{\href{http://astrothesaurus.org/uat/322}{Cosmic microwave background radiation (322)};  \href{http://astrothesaurus.org/uat/435}{Early Universe (435)}; \href{http://astrothesaurus.org/uat/1146}{Observational Cosmology (1146)}; \href{http://astrothesaurus.org/uat/1127}; \href{http://astrothesaurus.org/uat/799}{Astronomical instrumentation (799)}; 
\href{http://astrothesaurus.org/uat/1127}{Polarimeters (1127)}. 
\newpage
}

\section{Introduction}
The cosmic microwave background (CMB) anisotropy, in intensity and linear polarization, has added excellent constraining power to modern cosmological models \citep{Netterfield2002, Kovac2002, Bennett2013, Planck2018i, Louis2017, Henning2018}. Standard cosmology and the $\Lambda$CDM model predict no significant primordial circular polarization, yet constraints on this prediction are sparse. Theoretical work has produced a variety of predictions of circularly polarized emission mechanisms. Synchrotron emission in galaxies is intrinsically elliptically polarized  \citep{Legg1968}. Magnetized relativistic plasma in supernovae remnants of Population III stars \citep{Tashiro} and in galaxy clusters \citep{Cooray} can convert E-mode polarization into circular polarization through Faraday conversion. Primordial magnetic fields can generate circular polarization in the CMB \citep{Giovanni2009, Giovanni2010, Zarei2010}. Circular polarization can also result from coupling of the Chern-Simons term to electrodynamics resulting in cosmic birefringence \citep{Carroll}. Other mechanisms for the production of circularly polarized emission are discussed in \citet{Ejlli2019}, \citet{Mohammadi2013}, \citet{Sawyer2015}, and \citet{Venumadhav2017}, with a review in \citet{KingLubin}. All predictions are still several orders-of-magnitude fainter than what is accessible with the sensitivity levels of current experiments. The brightest predicted source of circularly polarized emission is the Zeeman transition in atmospheric Oxygen \citep{Keating, hanany}.

Existing constraints on circular polarization from extraterrestrial sources, reported by the \textsc{Spider} balloon-borne experiment \citep{spider}, are three-orders-of-magnitude larger than current linear polarization measurements \citep{Bicep2018}. The \textsc{Spider} measurement leveraged nonidealities in its half-wave plate polarization modulators, which provide a small amount of coupling to circular polarization, to produce an upper limit at 95 and 150\,GHz.

The Cosmology Large Angular Scale Surveyor (CLASS) is an array of four polarimetric telescopes situated in the Atacama Desert in Chile, operating at four frequencies spanning the microwave foreground minimum and targeting the polarized CMB at the largest angular scales \citep{Eimer2012, Essinger2014, Harrington2016}. Making measurements of the sky at these scales will provide powerful constraints on the optical depth to the epoch of reionization \citep{WattsE} and cosmological inflationary models \citep{WattsB}. The Variable-delay Polarization Modulator (VPM) technology employed by CLASS to access the largest scales is unique in that it provides explicit sensitivity to circular polarization (Section \ref{sec-vsensitivity}) \citep{Chuss2012, Harrington2018}. CLASS has been observing 75\% of the sky with the 40\,GHz instrument since 2016, and the remaining frequencies of 90, 150, and 220\,GHz have been deployed and are currently collecting data. We report on the circular polarization results for the first two years of observation with the 40\,GHz instrument over 56\% of the sky.

Section \ref{sec-data} describes the data selection and analysis, including an explanation of the instrument's coupling to circular polarization. Section \ref{sec-dipole} reports on a first detection of atmospheric emission at 40\,GHz and discusses its implications for CMB measurements. Section \ref{sec-upperlim} presents new upper limit constraints on circularly polarized emission at 40\,GHz.

\section{Data processing}
\label{sec-data}
\subsection{Sensitivity to Circular Polarization}
\label{sec-vsensitivity}
\EPSFIG[scale=0.7]{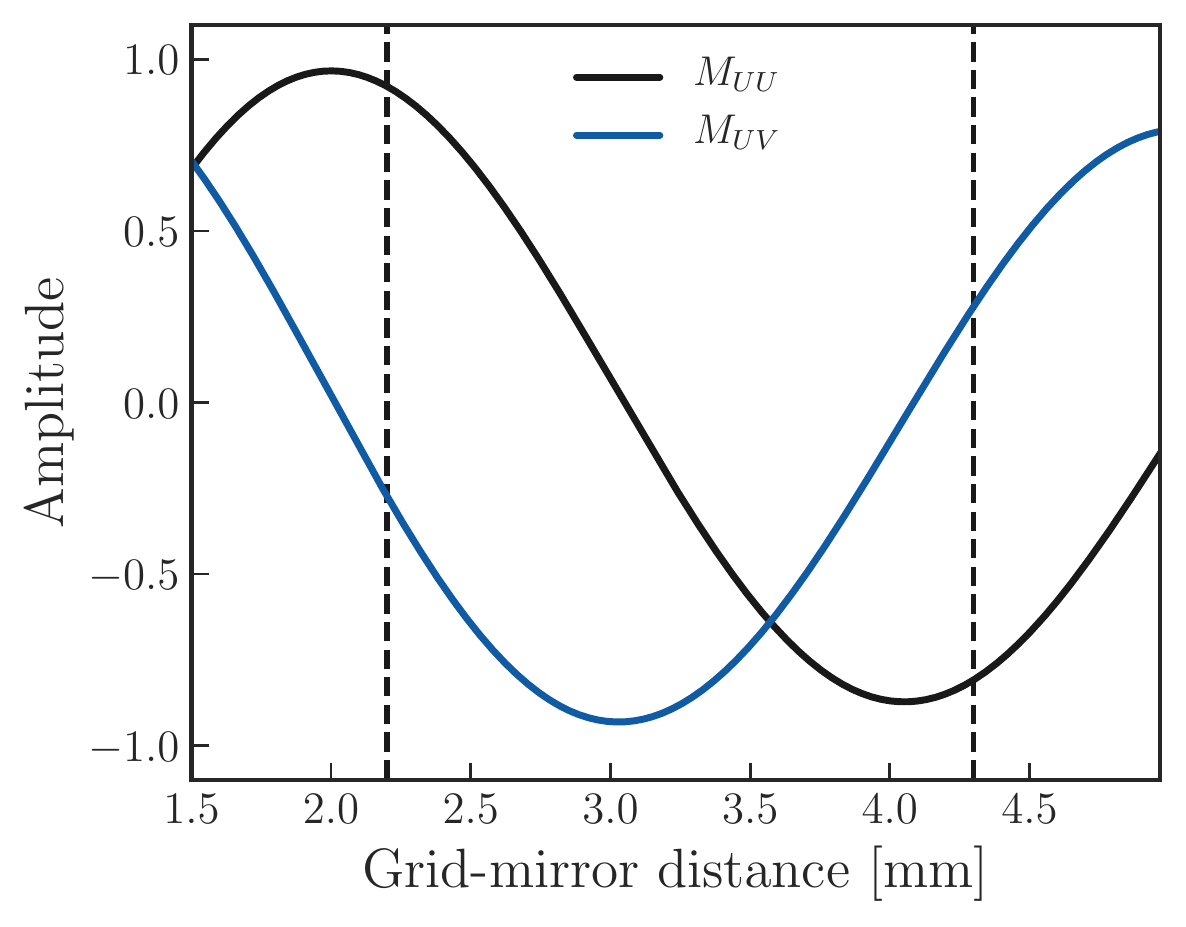}{VPM modulation transfer functions for incident $U$ and $V$ polarization. The dashed vertical lines indicate the throw of the VPM for the 40\,GHz instrument, through which Stokes $U$ modulates into $-U$ through the Stokes $V$.}{FIG-modfun}{t}

The CLASS front-end, fast modulation is provided by a Variable-Delay Polarization Modulator described in detail in  \citet{Chuss2012}, and \citet{Harrington2018}. The VPM consists of a stationary grid of parallel wires that reflects light polarized along one axis and transmits light polarized along the other. The transmitted light reflects off a flat mirror behind the grid and recombines with light reflected by the grid with an optical path difference that depends on the grid-mirror separation. The mirror is actuated to modulate the optical path difference at 10\,Hz, moving the signal band to higher frequencies that are less vulnerable to $1/f$ and other low-frequency noise \citep{Miller2016}. During one cycle of operation, the VPM switches the Stokes parameter $U$ for the incoming light into Stokes $V$, and then into Stokes $-U$ (\autoref{FIG-modfun}).

This explicit sensitivity to circular polarization is so far unique to CMB experiments using Variable-delay Polarization Modulators such as CLASS \citep{Harrington2016}, and PIPER \citep{Gandilo2017}.

\subsection{Mapmaking and Data Processing}
\label{sec-mapmaking}

The first era of CLASS observations ran from September 2016 through February 2018. During this period, the 40\,GHz instrument collected data year-round at a rate of 200\,Hz with 64 polarization-sensitive bolometers \citep{Appel2014, Appel2019}. The telescope scans the sky by performing $720^\circ$ azimuth ``sweeps'' in alternating directions, at a constant rate of one degree per second, held at an elevation of $45^\circ$. Each day the instrument observes at one of seven different boresight orientations, which range from $-45^\circ$ to $+45^\circ$ in increments of $15^\circ$. For this analysis we select data from the best 239 days of observation based on their noise properties, representing 62\% of the total data collected. We discard data from detectors that performed poorly during the season, leaving a total of 57 bolometers. To identify and eliminate contamination of the data, we compute the variance of the timestream on windows of various lengths and apply cuts to regions with anomalous values. We apply an additional cut when the distance between a detector pointing and the Sun (Moon) is smaller than $20^\circ$ ($10^\circ$). The timestream calibration to thermal units is based on observations of the Moon and described in \citet{Appel2019}. In the last pre-processing step we apply deconvolutions that correct for the readout filter \citep{mce} and detector time constants. 

The space between the grid and the mirror creates a resonant cavity that produces a signal that is a function of the grid-mirror separation, and the differential temperature and emissivity of the grid wires and the sky \citep{Miller2016}. To remove this VPM-synchronous signal we subtract a time-domain template created by averaging the data in each sweep in bins of grid position, as given by linear encoders on the VPM.

A model for the timestream data, $d$, obtained by multiplying the Mueller matrix of all elements in the optics chain, is
\begin{equation}
d = I + M_{UU }Q \cos (2\gamma) + M_{UU} U \sin (2\gamma) + M_{UV}V,
\label{eq-datamodel}
\end{equation}
where $\gamma$ is the sum of the parallactic angle of the grid and telescope boresight angle, and $M_{ij}$ for $i,j \in \{I,Q,U,V\}$ are the elements of the $4\times4$ VPM Mueller matrix \citep{Miller2016}. These elements are a function of the grid-mirror separation $z$, $M_{ij} = M_{ij}(z)$. To produce a map, we invert this timestream model using the traditional linear least squares mapmaking equation
\begin{equation}
\mathbf{s} = (A^TN^{-1}A)^{-1}A^TN^{-1}d,
\label{EQ-mapmaking}
\end{equation}
where $\mathbf{s}$ is an $\mathrm{N_{pixels}}\times \mathrm{N_{Stokes}}$ vector representing the sky maps of each Stokes parameter, $A$ is the projection operator mapping the data between the time-domain and the map-domain, and $N$ is the noise covariance matrix in the time basis. For this work, we employ a naive mapmaking approach where we approximate $N$ as a diagonal matrix with elements equal to the variance of the time segment being mapped, equivalent to one sweep of the telescope. To make this approximation valid for the Stokes $I$ component, we suppress the temperature signal from the atmosphere, which has a $1/f$ spectrum and is correlated between detectors. We do this by fitting an order 19 polynomial in the azimuth-domain to each timestream and subtracting it. While this biases the large-scale modes in the temperature maps, the fast modulation shifts the linearly and circularly polarized signals up in the frequency-domain to a band centered at 10\,Hz, where their amplitudes are not attenuated by this filtering. The recovered $V$ map is nearly unbiased (\autoref{FIG-transffunc}). We have extended the \texttt{Ninkasi} mapmaking code written for the \textsc{ACT} experiment \citep{dunner} to incorporate \autoref{eq-datamodel}. We use it  to perform all the data operations described in this section and to bin the maps into a \texttt{HealPix} grid in equatorial coordinates (\autoref{FIG-stokesv}). In this analysis we consider only the Stokes $V$ maps. We use the temperature maps (Stokes $I$) for calibration of the instrument \citep{Appel2019}. The Stokes $Q$ and $U$ maps will serve as a consistency check to the maps produced by the main linear polarization analysis pipeline which employs a destriping mapmaker (J. Eimer et al. (2019, in preparation)).

Before analyzing each map we discard the edges of the observing region by only keeping pixels visited by every detector, bringing the covered sky fraction to 56\%. We perform the analysis on the region of the sphere between declinations of $-60^\circ$\,S and $15^\circ$\,N. No galactic mask is applied. 

Since CLASS observes at fixed elevation, the circularly polarized atmospheric signal depends only on azimuth. With the telescope's observing strategy, this time-invariant azimuthal dependence implies a signal in the maps that is approximately independent of right ascension. To estimate the power spectrum of the extraterrestrial sky from the maps, we discard modes contaminated by azimuth-synchronous signals by subtracting the mean of the map at each declination. In the next section we describe how we estimate the power spectrum of the maps, taking into account the effects of the time- and map-domain filtering.


\subsection{Power Spectrum Estimation}

The selected data set consists of 239 constant elevation scans (CES) with the 40\,GHz telescope, each between 10 and 24 hours in duration. The average length of one CES is 20 hours, yielding a total of 4772 hours of observation including night and daytime data. We make 26 independent maps with the data distributed so that each has full coverage of the observation region and similar amounts of integration time. When distributing the data between the maps, we prioritize spatial coverage and map weight over boresight rotation coverage, since, unlike linear polarization, circular polarization measurements do not require observations at different telescope boresight orientations.

From these maps we produce $26 \choose 2$\,$=325$ cross-spectra using the \texttt{PolSpice} estimator described in \citet{polspice}. We use ensembles of simulations of an isotropic Gaussian random field with a flat spectrum to check the robustness of the estimator given our sky mask. These results are insensitive to other choices of continuous input spectrum. We use these simulations to verify that  \texttt{PolSpice} returns unbiased bandpowers in bins with width $\Delta \ell = \binwidth$ starting at $\ell=1$ for our choice of sky mask. Based on these simulations, we choose to forego the apodization of the correlation function. Since $V$ is a Spin-0 field, the treatment follows that of a temperature map.

When computing the power spectra, we weight each map by the $VV$ element of the naive pixel covariance matrix, which is an estimate of the noise in each pixel accounting for integration time and the modulation efficiency of the VPM in $V$. This map-domain weighting accounts for the inhomogeneity of the noise in the maps and reduces the variance across all bins by $\sim$15\%.

\EPSFIG[scale=0.58]{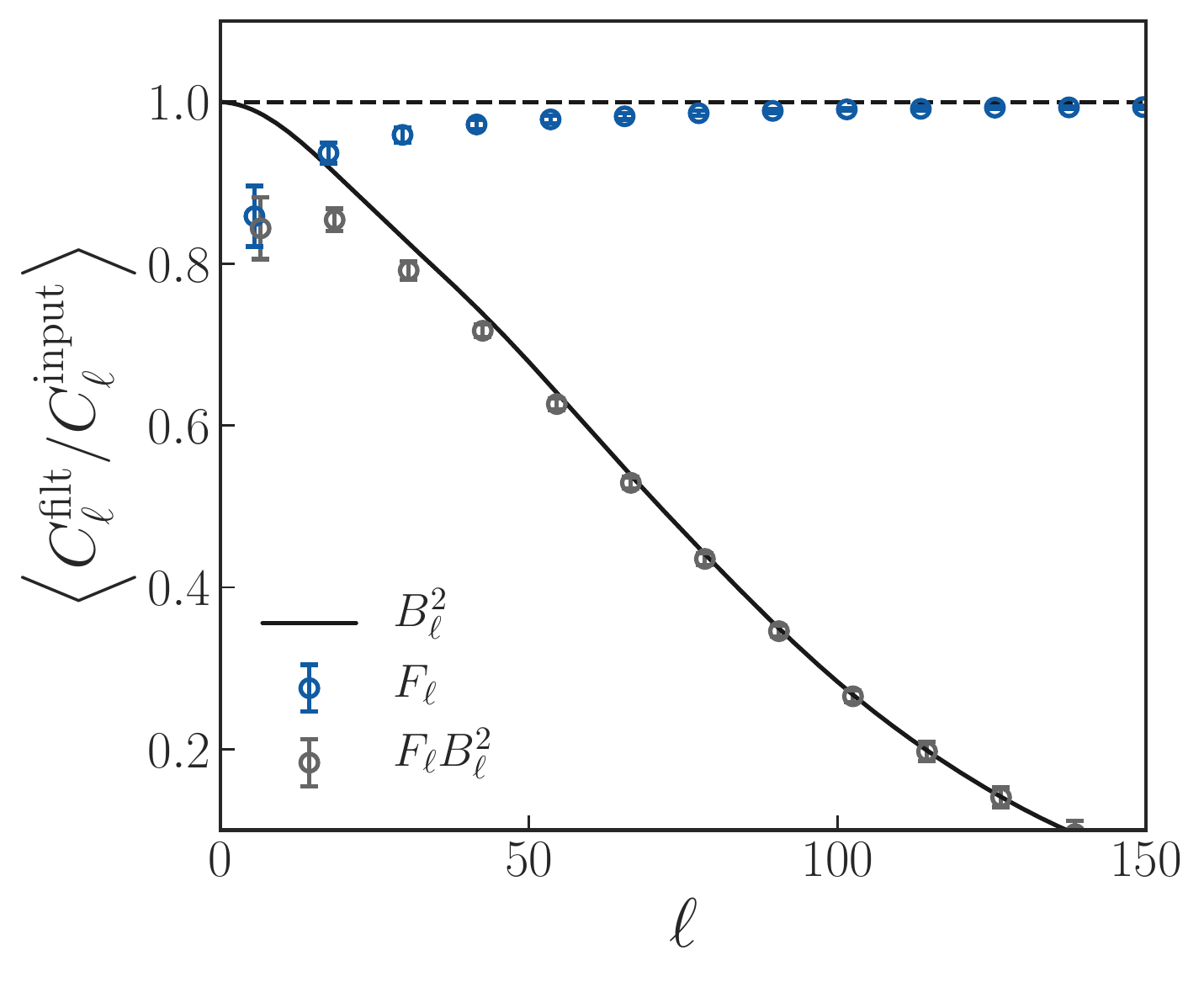}{Mapmaking $VV$ transfer function is shown in blue. The beam window function corresponding to the instrument's 1.5\,\degree\,FWHM beam is shown as a solid black line. These two quantities determine the experiment's sensitivity as a function of multipole. The gray points are the power spectrum window function $W_\ell = F_\ell B_\ell^2$. In this figure the filter transfer function points are artificially offset by $\Delta\ell=-1$ to aid visualization. }{FIG-transffunc}{}

To characterize the effects of the time- and map-domain filtering, we process a simulation ensemble through our pipeline. These simulations use the same data selection, observing times, and filtering used to make the data maps. The $m$-averaged filter transfer function, $F_\ell$, quantifying the amount of signal attenuated by the time- and map-domain filtering (described in Section \ref{sec-mapmaking}) is given by the mean of the ratios of the power spectra of the filtered simulated maps to the input simulated maps. The uncertainty in each bin of the transfer function is given by the spread of the distribution of the simulations. The gain of the instrument as a function of multipole is given by the instrument's beam window function, $B_\ell^2$. The beam window function is based on observations of the Moon, and described in detail in Z. Xu et al. (2019, in preparation). The product of these two quantities is the power spectrum window function, $W_\ell = F_\ell B_\ell^2$, used to debias the measured bandpowers. We bin this window function in bins with width $\Delta\ell=\binwidth$, the same as the data, to reduce correlations between adjacent multipoles at low $\ell$.

Note from \autoref{FIG-transffunc} that the filtering attenuates at most 14\% of the power, in the lowest bin, which spans ${1\leq\ell\leq12}$. 

\section{Results}
\label{sec-results}

We present a new upper limit to extraterrestrial circular polarization emission at $ 1 \leq \ell \leq 120$, and the detection of circularly polarized atmospheric  emission, both at 40\,GHz. The characteristics of the atmospheric emission in particular are of interest to current and future ground-based experiments targeting high precision polarization measurements at large angular scales. 

We check for self-consistency of the maps with a permutation test. From the 26 maps we construct 13 difference maps by randomly drawing pair combinations. We compute all possible cross-spectra from these maps and average them to estimate the null spectrum, and then compute the $p$-value of this null spectrum from the sample covariance matrix of all the cross-spectra. We repeat this operation 500 times. The distribution of these 500 $p$-values is uniformly distributed between zero and one, indicating that all observed signal is common to all maps. This rules out statistically significant amounts of time-varying systematic errors that might affect the maps unevenly. A more comprehensive analysis of systematic errors investigating other splits of the data is left for future work.

\subsection{Atmospheric Dipole}
\label{sec-dipole}

\EPSFIG[scale=0.7]{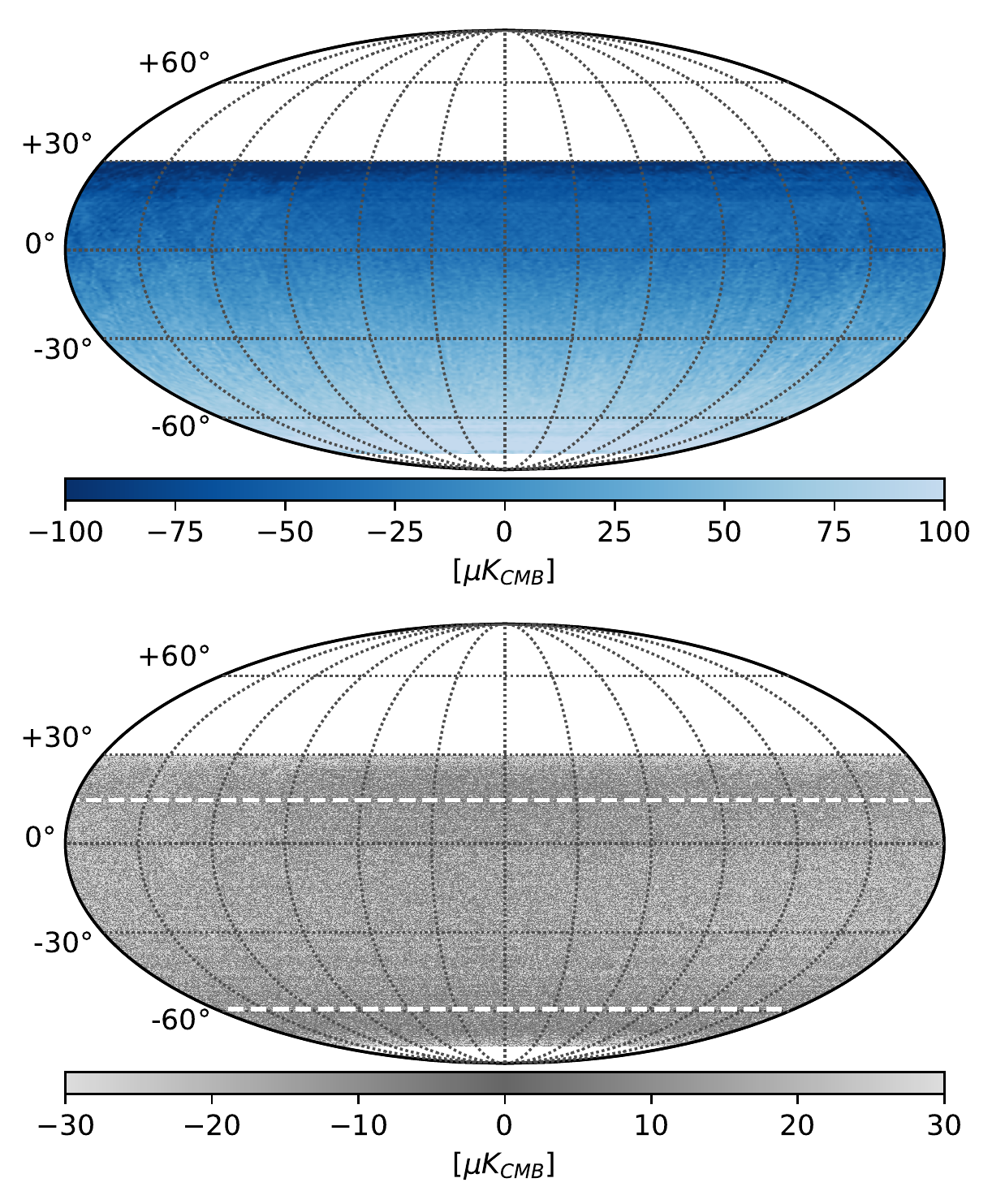}{(Top) Stokes $V$ map made from 239 days of CLASS data. This map is coadded from the 26 maps used in the cross-spectra analysis of Section \ref{sec-upperlim}. The gradient-like dipole structure visible in the map is caused by Zeeman emission of atmospheric oxygen, which depends on the inner product of the telescope's boresight vector and the Earth's magnetic field vector. This creates a pattern that is approximately only a function of declination. (Bottom) Stokes $V$ map coadded from declination-filtered maps used in the upper limit analysis. The solid white regions are not observed by the CLASS survey. The area between the white dashed lines is the analysis region. }{FIG-stokesv}{}

The main component seen in the maps is the polarized atmospheric emission (\autoref{FIG-stokesv}), caused by the Zeeman transitions of oxygen in the atmosphere as predicted by \citet{Keating}, \citet{hanany}, and \citet{Spinelli2011}. The Earth's magnetic field causes a splitting of the energy levels of oxygen in the atmosphere. Transitions between these energy levels are polarized, but have weak to no dependence on time-varying environmental parameters such as precipitable water vapor, temperature, and atmospheric pressure.

In the 40\,GHz $V$ map the majority of the atmospheric component is well modeled by a dipole with an amplitude of $\dipoleamplitude\pm\dipoleerror$\,$\mathrm{\mu K}$. This corresponds to a spherical harmonic mode at $(\ell,m)=(1,0)$ with coefficient $a_{\ell m}$ equal to $a_{10}=255\pm8\,\mathrm{\mu K}$. A detailed analysis of this component, including physically motivated atmospheric modeling, is discussed in M. Petroff et al. (2019, in preparation).

\EPSFIG[scale=0.58]{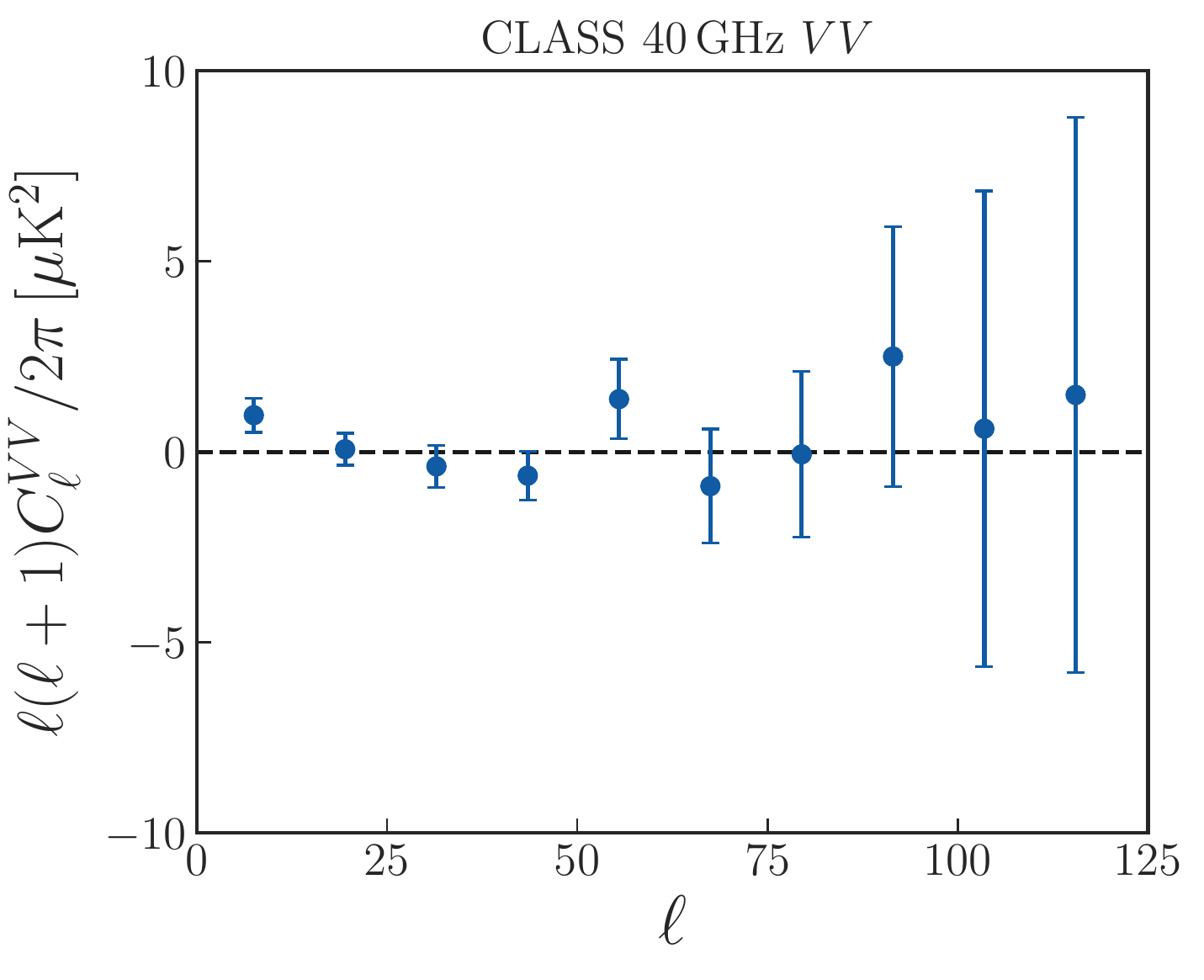}{$VV$ power spectrum at 40\,GHz after suppression of atmospheric component, binned into bins with $\Delta \ell=\binwidth$. The data has a reduced-$\chi^2=0.921$. Monte Carlo simulations made with the sample bin-bin covariance matrix yield a $p$-value of \pvalue for the data, indicating no statistically significant detection of power.}{FIG-vspectrum}{}
\begin{figure*}[t]               
\begin{center}                  %
\includegraphics[width=400pt, scale=0.4]{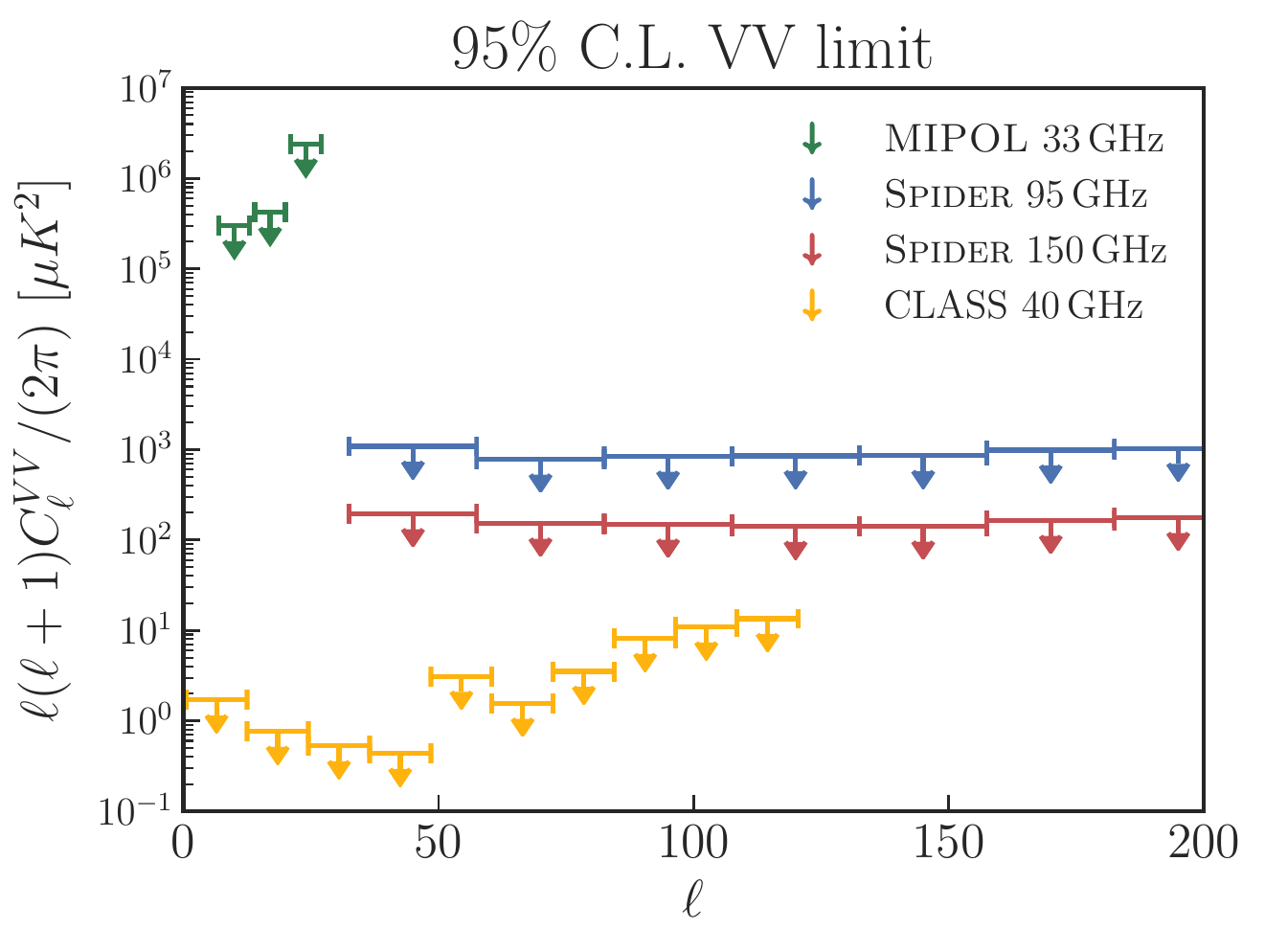}        %
\end{center}                    %
\caption{CLASS 95\% C.L. upper limits to extraterrestrial circular polarization at 40 GHz. The \textsc{MIPOL} \citep{mipol} limits at 33 GHz, and the \textsc{Spider} limits \citep{spider} at 95 GHz and 150 GHz, are shown for comparison in the $1 \leq \ell \leq 200$ range. The CLASS bins are $\Delta \ell=\binwidth$ wide starting at $\ell=1$, and going to $\ell=120$.}                    %
\label{FIG-upperlim}                      %
\end{figure*}   
With the contaminated declination-dependent modes suppressed by the declination filter we are left with an estimate of the circular polarization coming from the extraterrestrial sky. We model the effects of the filter statistically with Monte Carlo simulations to debias the $VV$ power spectrum, shown in \autoref{FIG-vspectrum}. A 95\% C.L. upper limit is constructed from this spectrum and is shown in \autoref{FIG-upperlim}.

The amplitude of the atmospheric emission is proportional to the inner product of the telescope's line-of-sight vector and the Earth's magnetic field vector \citep{Lenoir}. A map-domain declination filter that subtracts the mean along each row of pixels, applied independently to each map in the splits, suppresses the signal below measurable levels (\autoref{FIG-vspectrum}). The declination-filtered maps are used to estimate the power spectrum of the extraterrestrial signal. 

The effectiveness of the declination filter indicates the detected atmospheric signal is confined to the small fraction of modes that are not a function of right ascension. Discarding these modes corresponds to a loss of sensitivity of $\sim$1\% at $\ell=80$, the location of the recombination peak in the B-mode spectrum (\autoref{FIG-transffunc}). The highly localized nature of the contaminated modes relaxes the tolerances for instrumental V$\rightarrow$P leakage for instruments targeting $r\lesssim 0.01$ from the ground.

\subsection{Upper Limit on Circular Polarization Emission}
\label{sec-upperlim}

The error bars plotted in \autoref{FIG-vspectrum} are the square root of the diagonal elements of the bin-bin sample covariance matrix for all of these cross-spectra, divided by the square root of the number of cross-spectra. Uncertainties from the filter transfer function and beam window function are propagated assuming Gaussianity of the distributions in each bin. This assumption is valid for all bins given the chosen binwidth. The data has a reduced-$\chi^2=0.921$, which has a $p$-value of \pvalue$\!\!$, indicating no statistically significant detection of power over the multipole range $1 \leq \ell \leq 120$. This $p$-value is estimated via Monte Carlo simulations with samples drawn from the sample covariance matrix.

From this power spectrum we produce an upper limit per bin equal to the estimated bandpower value plus a 95\% confidence limit, shown in \autoref{FIG-upperlim}. This upper limit is a five-orders-of-magnitude improvement over the previous limit at Ka band by the \textsc{MIPOL} experiment at large scales \citep{mipol}, and a two-orders-of-magnitude improvement over the previously best CMB upper limits by \textsc{Spider} \citep{spider}.

\section{Conclusion}

We present circular polarization results from the first two years of observation with the CLASS 40\,GHz instrument, consisting of a detection of atmospheric emission and an upper limit on extraterrestrial emission in the $1 \leq \ell \leq 120$ range. The atmospheric signal, when measured at $45^\circ$ elevation, has an amplitude of $\dipoleamplitude\pm\dipoleerror$\,$\mathrm{\mu K}$. The circular polarization upper limit at 40\,GHz ranges from $\uplimmin\,\mathrm{\mu K}^2$ to $\uplimmax\,\mathrm{\mu K}^2$ (\autoref{FIG-upperlim}), representing a two-orders-of-magnitude better constraint on extraterrestrial circular polarization emission, and a five-orders-of-magnitude improvement at the largest scales.

CLASS telescopes covering three additional frequencies, 90, 150, and 220\,GHz, are currently equipped with VPMs and operating in the Atacama Desert in Chile. These instruments will provide improved frequency coverage of circular polarization emission, which will help characterize the atmospheric emission at frequencies relevant to current and future CMB polarization experiments, and add considerable constraining power to galactic, extragalactic, and cosmological sources of circularly polarized emission. Higher frequency CLASS data currently being collected will improve the multipole coverage thanks to the increased resolution of the higher frequency instruments.

\section*{Acknowledgments}
We acknowledge the National Science Foundation Division of Astronomical Sciences for their support of CLASS under Grant Numbers 0959349, 1429236, 1636634, and 1654494.
We thank Johns Hopkins University President R. Daniels and Dean B. Wendland for their steadfast support of CLASS.
We further acknowledge the very generous support of Jim and Heather Murren (JHU A\&S '88), Matthew Polk (JHU A\&S Physics BS '71), David Nicholson, and Michael Bloomberg (JHU Engineering '64).
The CLASS project employs detector technology developed in collaboration between JHU and Goddard Space Flight Center under several previous and ongoing NASA grants. Detector development work at JHU was funded by NASA grant number NNX14AB76A.
R.R. acknowledges partial support from CATA, BASAL grant AFB-170002, and Conicyt-FONDECYT through grant 1181620. R.D. thanks CONICYT for grant BASAL CATA AFB-170002.
We acknowledge scientific and engineering contributions from Max Abitbol, Fletcher Boone, Francisco Espinoza, Connor Henley, Lindsay Lowry, Isu Ravi, Gary Rhodes, Bingie Wang, Zi'ang Yan, Qinan Wang, and Tiffany Wei. 
We acknowledge productive collaboration with Dean Carpenter and the JHU Physical Sciences Machine Shop team.
Part of this research project was conducted using computational resources at the Maryland Advanced Research Computing Center (MARCC).
Part of this research project was conducted using computational resources of the Geryon-2 cluster at Centro de Astroingenier\'ia UC.
CLASS is located in the Parque Astron\'omico Atacama in northern Chile under the auspices of the Comisi\'on Nacional de Investigaci\'on Cient\'ifica y Tecnol\'ogica de Chile (CONICYT).

\software{\texttt{IPython} \citep{ipython}, \texttt{numpy} \citep{numpy}, \texttt{scipy} \citep{scipy}, \texttt{matplotlib} \citep{matplotlib}, \texttt{HEALPix} \citep{Gorski2005, healpy}, \texttt{CAMB} \citep{CAMB}, \texttt{GNU Octave} \citep{octave}.}


\begin{thebibliography}{}
\bibitem[Appel et al.(2014)]{Appel2014} Appel, J.~W., Ali, A., Amiri, M., et al.\ 2014, \procspie, 91531J
\bibitem[Appel et al.(2019)]{Appel2019} Appel, J.~W., Xu, Z., Padilla, I.~L., et al.\ 2019, \apj, 876, 126

\bibitem[Battistelli et al.(2008)]{mce} Battistelli, E.~S., Amiri, M., Burger, B., et al.\ 2008, Journal of Low Temperature Physics, 151, 908

\bibitem[Bennett et al.(2013)]{Bennett2013} Bennett, C.~L., Larson, D., Weiland, J.~L., et al.\ 2013, \apjs, 208, 20
\bibitem[BICEP2 Collaboration et al.(2018)]{Bicep2018} BICEP2 Collaboration, Keck Array Collaboration, Ade, P. A. R., et al.\ 2018, \prl, 121, 221301

\bibitem[Carroll et al.(1990)]{Carroll} Carroll, S.~M., Field, G.~B., \& Jackiw, R.\ 1990, \prd, 41, 1231
\bibitem[Chon et al.(2004)]{polspice} Chon, G., Challinor, A., Prunet, S., et al.\ 2004, \mnras, 350, 914
\bibitem[Chuss et al.(2012)]{Chuss2012} Chuss, D.~T., Wollack, E.~J., Henry, R., et al.\ 2012, \ao, 51, 197
\bibitem[Cooray et al.(2003)]{Cooray} Cooray, A., Melchiorri, A., \& Silk, J.\ 2003, Physics Letters B, 554, 1

\bibitem[De \& Tashiro(2015)]{Tashiro} De, S., \& Tashiro, H.\ 2015, \prd, 92, 123506
\bibitem[D{\"u}nner et al.(2013)]{dunner} D{\"u}nner, R., Hasselfield, M., Marriage, T.~A., et al.\ 2013, \apj, 762, 10

\bibitem[John W. Eaton et al.(2019)]{octave} Eaton, J.~W., Bateman, D., Hauberg, S., Wehbring, R. \ 2019 \ \texttt{GNU Octave} version 4.2.0

\bibitem[Eimer et al.(2012)]{Eimer2012} Eimer, J.~R., Bennett, C.~L., Chuss, D.~T., et al.\ 2012, \procspie, 845220
\bibitem[Ejlli(2019)]{Ejlli2019} Ejlli, D.\ 2019, European Physical Journal C, 79, 231

\bibitem[Essinger-Hileman et al.(2014)]{Essinger2014} Essinger-Hileman, T., Ali, A., Amiri, M., et al.\ 2014, \procspie, 91531I
\bibitem[Errard et al.(2015)]{Errard2015} Errard, J., Ade, P.~A.~R., Akiba, Y., et al.\ 2015, \apj, 809, 63

\bibitem[Gandilo et al.(2017)]{Gandilo2017} Gandilo, N., Ade, P., Benford, D.~J., et al.\ 2017, American Astronomical Society Meeting Abstracts \#229 229, 430.04

\bibitem[Giovannini(2009)]{Giovanni2009} Giovannini, M.\ 2009, \prd, 80, 123013
\bibitem[Giovannini(2010)]{Giovanni2010} Giovannini, M.\ 2010, \prd, 81, 023003
\bibitem[G{\'o}rski et al.(2005)]{Gorski2005} G{\'o}rski, K.~M., Hivon, E., Banday, A.~J., et al.\ 2005, \apj, 622, 759


\bibitem[Hanany \& Rosenkranz(2003)]{hanany} Hanany, S., \& Rosenkranz, P.\ 2003, \nar, 47, 1159
\bibitem[Harrington et al.(2016)]{Harrington2016} Harrington, K., Marriage, T., Ali, A., et al.\ 2016, \procspie, 99141K
\bibitem[Harrington et al.(2018)]{Harrington2018} Harrington, K., Eimer, J., Chuss, D.~T., et al.\ 2018, \procspie, 107082M
\bibitem[Henning et al.(2018)]{Henning2018} Henning, J.~W., Sayre, J.~T., Reichardt, C.~L., et al.\ 2018, \apj, 852, 97
\bibitem[Hunter(2007)]{matplotlib} Hunter, J.~D.\ 2007, Computing in Science and Engineering, 9, 90

\bibitem[Keating et al.(1998)]{Keating} Keating, B., Timbie, P., Polnarev, A., et al.\ 1998, \apj, 495, 580
\bibitem[King \& Lubin(2016)]{KingLubin} King, S., \& Lubin, P.\ 2016, \prd, 94, 023501
\bibitem[Kovac et al.(2002)]{Kovac2002} Kovac, J.~M., Leitch, E.~M., Pryke, C., et al.\ 2002, \nat, 420, 772

\bibitem[Lange et al.(2001)]{Lange2001} Lange, A.~E., Ade, P.~A., Bock, J.~J., et al.\ 2001, \prd, 63, 042001
\bibitem[Legg \& Westfold(1968)]{Legg1968} Legg, M.~P.~C., \& Westfold, K.~C.\ 1968, \apj, 154, 499
\bibitem[Lenoir(1968)]{Lenoir} Lenoir, W.~B.\ 1968, \jgr, 73, 361
\bibitem[Lewis et al.(2000)]{CAMB} Lewis, A., Challinor, A., \& Lasenby, A.\ 2000, \apj, 538, 473
\bibitem[Louis et al.(2017)]{Louis2017} Louis, T., Grace, E., Hasselfield, M., et al.\ 2017, \jcap, 2017, 031

\bibitem[Mainini et al.(2013)]{mipol} Mainini, R., Minelli, D., Gervasi, M., et al.\ 2013, \jcap, 2013, 033
\bibitem[Miller et al.(2016)]{Miller2016} Miller, N.~J., Chuss, D.~T., Marriage, T.~A., et al.\ 2016, \apj, 818, 151
\bibitem[Mohammadi(2013)]{Mohammadi2013} Mohammadi, R.\ 2013, arXiv e-prints, arXiv:1312.2199

\bibitem[Nagy et al.(2017)]{spider} Nagy, J.~M., Ade, P.~A.~R., Amiri, M., et al.\ 2017, \apj, 844, 151
\bibitem[Netterfield et al.(2002)]{Netterfield2002} Netterfield, C.~B., Ade, P.~A.~R., Bock, J.~J., et al.\ 2002, \apj, 571, 604
\bibitem[Perez \& Granger(2007)]{ipython} Perez, F., \& Granger, B.~E.\ 2007, Computing in Science and Engineering, 9, 21

\bibitem[Planck Collaboration et al.(2018)]{Planck2018i} Planck Collaboration, Akrami, Y., Arroja, F., et al.\ 2018, arXiv e-prints, arXiv:1807.06205

\bibitem[Sawyer(2015)]{Sawyer2015} Sawyer, R.~F.\ 2015, \prd, 91, 021301
\bibitem[Spinelli et al.(2011)]{Spinelli2011} Spinelli, S., Fabbian, G., Tartari, A., et al.\ 2011, \mnras, 414, 3272

\bibitem[van der Walt et al.(2011)]{numpy} van der Walt, S., Colbert, S.~C., \& Varoquaux, G.\ 2011, Computing in Science and Engineering, 13, 22

\bibitem[Venumadhav et al.(2017)]{Venumadhav2017} Venumadhav, T., Oklop{\v{c}}i{\'c}, A., Gluscevic, V., et al.\ 2017, \prd, 95, 083010
\bibitem[Virtanen et al.(2019)]{scipy} Virtanen, P., Gommers, R., Oliphant, T.~E., et al.\ 2019, arXiv e-prints, arXiv:1907.10121

\bibitem[Watts et al.(2015)]{WattsB} Watts, D.~J., Larson, D., Marriage, T.~A., et al.\ 2015, \apj, 814, 103
\bibitem[Watts et al.(2018)]{WattsE} Watts, D.~J., Wang, B., Ali, A., et al.\ 2018, \apj, 863, 121


\bibitem[Zarei et al.(2010)]{Zarei2010} Zarei, M., Bavarsad, E., Haghighat, M., et al.\ 2010, \prd, 81, 084035
\bibitem[Zonca et al.(2019)]{healpy} Zonca, A., Singer, L., Lenz, D., et al.\ 2019, The Journal of Open Source Software, 4, 1298
\end{thebibliography}
\end{document}